# Classical Electron Model with QED Corrections

Ron Lenk*

*Marietta, GA, USA*

June 8, 2010

**Abstract**

In this article we build a metric for a classical general relativistic electron model with QED corrections. We calculate the stress-energy tensor for the radiative corrections to the Coulomb potential in both the near-field and far-field approximations. We solve the three field equations in both cases by using a perturbative expansion to first order in $\alpha$ (the fine-structure constant) while insisting that the usual (+, +, -, -) structure of the stress-energy tensor is maintained. The resulting metric models a (non-spinning) electron with a Coulomb potential with QED corrections, and maintains masslessness of the photon to self-consistent order. The near-field solution resembles the metric of a global monopole.

PACS number(s): 12.20-m, 04.25.Nx, 04.40.Nr, 04.70.Dy

Keywords: electron model; radiative corrections; global monopole

As is well-known, Birkhoff's Theorem establishes that the only static, spherically symmetric, asymptotically flat metric is the Reissner-Nordström metric [1]. Since its stress-energy tensor has zero trace and the structure (+, +, -, -), with the individual terms dropping off as $r^{-4}$, it can be considered as a first-approximation to a classical electron model [2]. Of course, it doesn't include spin, and for actual electron parameters it exhibits a naked singularity, and is thus not wholly satisfactory.

The goal of this paper is to increase the resemblance of the Reissner-Nordström metric to an electron by adding in terms that are due to vacuum polarization. According to QED [3], the presence of virtual electron-positron pairs in the strong field near to an electron modify the classical $q$ / $r$ Coulomb potential because the virtual positrons are preferentially attracted to the electron while the virtual electrons are repelled. The result is that terms logarithmic in $r$ (in the near-field) or exponential in $-r$ (in the far-field) appear in the potential. Since the exact solution to the equations isn't known, it is traditional to approximate the potential by a series in $\alpha$, the fine-structure constant. Since $\alpha \approx 1 / 137$, this series converges rapidly. In this series expansion, additional terms proportional to $q / r$ appear, and these are absorbed into the classical Coulomb potential by renormalization, which is to say redefining $q$.

* e-mail: ron.lenk@reliabulb.com

In this paper we follow similar reasoning. The corrected near-field Coulomb potential to first-order in $\alpha$ is given by [3]

$$\Phi(r) = \frac{q}{r}\left[1 + \frac{2\alpha}{3\pi}\left(\ln\left(\frac{1}{mr}\right) - \gamma - \frac{5}{6}\right)\right] \tag{1a}$$

with $\gamma$ the Euler-Mascheroni constant = 0.5772…. This expression is valid in the range where $r << \hbar / mc = 4 \times 10^{-16} m$. In the opposite regime, where $r >> 4 \times 10^{-16} m$, the corrected far-field potential is given by

$$\Phi(r) = \frac{q}{r}\left(1 + \frac{\alpha}{4\sqrt{\pi}} \frac{e^{-2mr}}{(mr)^{3/2}}\right) \tag{1b}$$

Now the $T^0{}_0$ component of the stress-energy tensor equals $E^2 = (-\nabla\Phi)^2$. For Eq. (1a) we get

$$T^0{}_0 = \frac{q^2}{r^4}\left[1 + \frac{\alpha}{3\pi}\left(4\ln\left(\frac{1}{mr}\right) - 4\gamma - \frac{10}{3}\right) + \alpha^2(\ldots)\right] \tag{2a}$$

and for Eq. (1b)

$$T^0{}_0 = \frac{q^2}{r^4}\left[1 + \frac{\alpha}{mr^{3/2}e^{2mr}\sqrt{\pi}}\left(\frac{1}{2} + mr\right) + \alpha^2(\ldots)\right] \tag{2b}$$

In these expressions, we have self-consistently removed the terms in $\alpha^2$, since Eqns. (1a) and (1b) are only valid to first order in $\alpha$. We will also renormalize the charge in Eqn. (1a) to absorb the terms proportional to $r^{-4}$, leaving

$$T^0{}_0 = \frac{q^2}{r^4}\left[1 - \frac{4\alpha}{3\pi}\ln(r)\right] \tag{3a}$$

and

$$T^0{}_0 = \frac{q^2}{r^4}\left[1 + \frac{\alpha}{mr^{3/2}e^{2mr}\sqrt{\pi}}\left(\frac{1}{2} + mr\right)\right] \tag{3b}$$

Now we will take an ansatz for the metric. Since Eqns. (1a) and (1b) represent small deviations from the classical Coulomb potential, we will assume that the metric has small deviations from the Reissner-Nordström form. Since we will also want to ensure the standard trace-free form of the stress-energy tensor

$$T^\mu_\nu = diag(+,+,-,-) \tag{4}$$

we will also give ourselves some latitude with a function modifying the $(\theta, \theta)$ and $(\phi, \phi)$ components. Our ansatz is

$$g_{\mu\nu} = \begin{pmatrix} A(r) & 0 & 0 & 0 \\ 0 & -(B(r))^{-1} & 0 & 0 \\ 0 & 0 & -r^2C(r) & 0 \\ 0 & 0 & 0 & -r^2\sin^2(\theta)C(r) \end{pmatrix} \tag{5}$$

with $A(r)$, $B(r)$ and $C(r)$ to be determined, and with $A(r)$ and $B(r)$ presumed to be of the form $1-\frac{2M}{r}+\frac{Q^2}{r^2}+f(\ln(r))$.

The equations, too lengthy to be reproduced here, can be solved using a symbolic algebra program (MAPLE) [4], and we get immediately that for the stress-energy tensor of Eq. (3a)

$$C(r)=\frac{C_a}{\sqrt{4\alpha\ln(r)-3\pi}} \tag{6a}$$

and for Eq. (3b)

$$C(r)=C_b-\frac{(1+2Mr)C_b\alpha}{4\sqrt{\pi}mr^{\frac{3}{2}}e^{2Mr}} \tag{6b}$$

with $C_a$ and $C_b$ integration constants.

These expressions can now be used to determine $A(r)$ and $B(r)$. The expressions for $B(r)$ involve an integral of $C(r)$ that cannot be directly evaluated. Instead, we expand them in powers of $\alpha$, retaining terms only to first order, consistent with the approximations in Eqns. (1). The expressions for $A(r)$ similarly involve an integral in both $C(r)$ and $B(r)$, and we expand them also to first order in $\alpha$.

We select the two integration constants in $B(r)$ and $C(r)$ to give the normal Reissner-Nordström form to $B(r)$. We also redefine the charge to incorporate $ln(r)$-free terms in $\alpha$ / $r^2$. The third integration constant, in $A(r)$, is selected to give one as the first term for $A(r)$. Taking $C(r)$ also to first order in $\alpha$, we end up with

$$A(r) = 1 - \frac{2M}{r} + \frac{8\pi Q^2}{r^2} + \frac{2\alpha \ln(r)}{3\pi}\left(\frac{M}{r} - \frac{8\pi Q^2}{r^2}\right)$$

$$B(r) = 1 - \frac{2M}{r} + \frac{8\pi Q^2}{r^2} + \frac{2\alpha \ln(r)}{3\pi}\left(1 + \frac{3M}{r} - \frac{16\pi Q^2}{r^2}\right) \tag{8a}$$

$$C(r) = \left(1 - \frac{2\alpha}{3\pi}\right) + \frac{2\alpha \ln(r)}{3\pi}$$

for the near-field and

$$A(r) = 1 - \frac{2M}{r} + \frac{8\pi Q^2}{r^2} + \frac{\alpha e^{-2Mr}}{4\sqrt{r^3 \pi}}\left(\frac{8\pi Q^2}{Mr^2} + \frac{16\pi Q^2}{r} - \frac{1}{r} - 2M\right)$$

$$B(r) = 1 - \frac{2M}{r} + \frac{8\pi Q^2}{r^2} + \frac{\alpha M e^{-2Mr}}{\sqrt{r\pi}}\left(-\frac{8\pi Q^2}{r} - r + 2M - \frac{1}{4M} + \frac{\pi Q^2}{m^2 r^3} + \frac{2\pi Q^2}{Mr^2} - \frac{1}{8M^2 r}\right) \tag{8b}$$

$$C(r) = 1 - \frac{(1 + 2Mr)\alpha}{4\sqrt{\pi} m r^{\frac{3}{2}} e^{2Mr}}$$

for the far-field. Our solutions have been obtained just by solving the individual equations of the terms of the stress-energy tensor. As a check, we substitute Eqns. (8) into the field equations and verify that the trace of the stress-energy tensor in both cases is indeed zero to order $\alpha$, as it should be.

The structure of the metric terms in Eqns. (8a) and (8b) is interesting. In both cases, the $\alpha^0$ terms for $A(r)$ and $B(r)$ match those of the Reissner-Nordström metric, even though the $M$ and $Q$ terms of $A(r)$ were not explicitly set. To the same order, $C(r)$ in Eq. (8b) is one, and so the metric has the same singularity structure as the Reissner-Nordström metric. However, $C(r)$ in Eq. (8a) is not one; it has a deficit spherical angle of $2\alpha / 3\pi \approx 0.0015$. Thus to lowest order, the far-field approximation of the QED corrected electron metric appears as a Reissner-Nordström black hole, but in near-field it is revealed rather as a global monopole.

Work is in progress to extend this solution to the Kerr-Newman metric, to see if there are QED-type corrections to the gyromagnetic ratio.